\RequirePackage{fix-cm}

\PassOptionsToPackage{dvipsnames,table}{xcolor}
\documentclass[smallextended]{svjour3}       
\smartqed  
\usepackage{graphicx}
\usepackage{natbib}
\usepackage{booktabs}
\usepackage{multirow} 
\usepackage{array}
\usepackage{listings}
\usepackage{float}
\usepackage{rotating}
\usepackage{verbatim}
\usepackage{url}
\usepackage{xcolor}
\usepackage{amsmath}
\usepackage{tabularx}
\usepackage{arydshln}
\usepackage{pifont}  

\usepackage{fix-cm}
\usepackage[T1]{fontenc}
\usepackage{lmodern}

\newcommand{\cmark}{\textcolor{green!55!black}{\ding{51}}}
\newcommand{\xmark}{\textcolor{red!70!black}{\ding{55}}}

\definecolor{codegreen}{rgb}{0,0.6,0}
\definecolor{codegray}{rgb}{0.5,0.5,0.5}
\definecolor{codepurple}{rgb}{0.58,0,0.82}
\definecolor{backcolour}{rgb}{0.95,0.95,0.92}

\usepackage[justification=centering]{caption}
\usepackage{titlesec}
\titlespacing*{\subsection}{0pt}{1em}{0.5em}
\begin{document}

\title{AI-Mediated Code Comment Improvement 
}


\author{Maria~Dhakal         \and
Chia-Yi~Su         \and
Robert~Wallace         \and
Chris~Fakhimi        \and
Aakash~Bansal         \and
Toby~Li         \and
Yu~Huang         \and
Collin~McMillan 
}


\institute{
        Maria~Dhakal \at
              Department of Computer Science and Engineering, University of Notre Dame, IN, USA. \\
              \email{mdhakal@nd.edu}           
\and
        Chia-Yi~Su \at
        Department of Computer Science and Engineering, University of Notre Dame, IN, USA.
        \and
Robert~Wallace        \at
        Department of Computer Science and Engineering, University of Notre Dame, IN, USA.
        \and
Chris~Fakhimi        \at
        Department of Computer Science and Engineering, University of Notre Dame, IN, USA.
        \and
Aakash~Bansal         \at
        Division of Computer Science and Engineering, Louisiana State University, LA, USA.
        \and
Toby~Li         \at
        Department of Computer Science and Engineering, University of Notre Dame, IN, USA.
        \and
Yu~Huang         \at
        Department of Computer Science, Vanderbilt University, Nashville, TN, USA
        \and
          Collin~McMillan \at
              Department of Computer Science and Engineering, University of Notre Dame, IN, USA.
}

\date{Received: date / Accepted: date}

\maketitle

\begin{abstract}
This paper describes an approach to improve code comments along different quality axes by rewriting those comments with customized Artificial Intelligence (AI)-based tools.  We conduct an empirical study followed by grounded theory qualitative analysis to determine the quality axes to improve.  Then we propose a procedure using a Large Language Model (LLM) to rewrite existing code comments along the quality axes.  We implement our procedure using GPT-4o, then distil the results into a smaller model capable of being run in-house, so users can maintain data custody.  We evaluate both our approach using GPT-4o and the distilled model versions.  We show in an evaluation how our procedure improves code comments along the quality axes.  We release all data and source code in an online repository for reproducibility.
\keywords{source code comments \and empirical studies \and grounded theory \and large language model \and code summarization}
\end{abstract}

\section{Introduction}\label{sec1}

Source code comments are a lynchpin of program comprehension and maintainability~\citep{song2019survey}.  But comments are notorious for being out of date, misleading, and generally of poor quality~\citep{hu2018deep}. Improving these comments is a valuable target for researchers, with entire research areas having formed variously called code summarization~\citep{mcburney2014automatic, ahmad2020transformer}, code comment generation~\citep{hu2018deep}, and niches such as commit message generation~\citep{zhang2024automatic}.  Yet a perennial problem in these research areas are the question ``what qualities make a code comment good?'' and ``how do we aim our tools to generate comments along those qualities?''\citep{song2019survey, rani2023decade}.  Almost all approaches aim to mimic code comments in repositories of human-written source code, but mimicking those comments is likely to produce new comments with the same problems that existed in the comments in the first place.

In this paper, we perform empirical work to understand how comments should improve, then we design and implement a procedure for improving comments.  For empirical work, we 1) conduct a study to compare source code comments and obtain programmers' rationales for why some comments are better than others, and 2) analyze the programmers' rationales using a grounded theory approach to find key ``quality axes'' along which code comments should be improved.  We then design an approach using a Large Language Model (LLM) to improve existing comments along the quality axes.  We implement the approach using both a commercial LLM (GPT-4o \citep{achiam2023gpt}) and a smaller model suitable for in-house execution (jam \citep{su2023language, su2024distilled}).  The in-house execution is important because it maintains data custody by not sending source code to a third party.

The niche this paper fills is aligning research on code comment quality with research on code comment generation. Code comment quality has been studied for years; for instance, \cite{khamis2010automatic} assessed the quality of inline comments based on consistency and language quality using a heuristic based approach. \cite{steidl2013quality} evaluated documentation comment quality across four quality attributes, such as consistency, coherence, completeness, and usefulness. \cite{zhou2017analyzing} proposed a heuristic and natural language processing-based technique to detect incomplete and incorrect comments.

Our study is different in that we focus specifically on \emph{improvements} to code comment quality.  We ask 10 programmers to read different code comments for the same code and decide on which they prefer.  Then we ask them to rewrite the code comment and give a rationale behind their decisions.  The result is a dataset of 10 subjective programmer opinion about code quality differences and improvements.  Our next step is to find axes of quality along which code comments should be improved.  We perform a grounded theory analysis using the Straussian approach~\citep{strauss1997grounded} to label the rationales given by programmers in our study.  We find seven ``quality axes'' that describe different aspects of improvements that programmers felt were helpful.

We propose an AI-mediated procedure for improving code comments along the seven quality axes we find.  Our procedure is essentially to: 1) extract samples of improvements along the quality axes to serve as training examples, 2) fine-tune an LLM-based text generator (e.g., GPT-4o) with the training examples, and 3) distill the output of the LLM into a smaller language model capable of being run in-house (e.g., jam).  Then we can use either the fine-tuned LLM or the distilled model to improve arbitrary source code comments along any of the quality axes.  Finally, perform an evaluation of our approach on a held out set of source code comments in a study with 70 programmers and show that improvements along the quality axes we propose lead to an overall improvement in perceived code comment quality.


\section{Background and Related Work}

This section reviews prior work across three key areas relevant to our study: (1) code comment quality, (2) techniques for code summarization and automated comment generation, and (3) the application of grounded theory in software engineering research, with a focus on how qualitative methodologies have been used to derive taxonomies, interpret developer practices, and inform tool design. Table~\ref{tab:screlated} lists recent works related to this paper, highlighting the diversity of papers selected across different domains for our study. 

\begin{table}[h]
\centering
	{\small
		\begin{tabular}{@{} l c c c @{}}
    \toprule
    \textbf{Literature}                 & \textbf{G} & \textbf{L} & \textbf{C} \\
    \midrule
    \cite{wen2019large}                 & \xmark     & \xmark     & \cmark     \\
    \cite{leclair2019neural}            & \xmark     & \xmark     & \cmark     \\
    \cite{ADOLPH20121269}               & \cmark     & \xmark     & \xmark     \\
    \cite{hoda2021socio}                & \cmark     & \xmark     & \xmark      \\
    \cite{chen2021my}                   & \xmark     & \xmark     & \cmark     \\
    \cite{roy2021reassessing}           & \cmark     & \xmark     & \xmark      \\
    \cite{rani2023decade}               & \xmark     & \xmark     & \cmark     \\
    \cite{lee2023rlaif}                 & \xmark     & \cmark     &  \xmark     \\
    \cite{su2024context}                & \xmark     & \cmark     & \cmark     \\
    \cite{zhou2024can}                  & \cmark     & \cmark     &  \xmark    \\
    \cite{geng2024large}                & \xmark     & \cmark     & \cmark     \\
    \cite{ahmed2024can}                 & \xmark     & \cmark     & \cmark     \\
    \cite{su2024semantic}               & \xmark     & \cmark     & \xmark     \\
    \cite{su2024distilled}              & \xmark     & \cmark     & \xmark     \\
    \midrule
    \textbf{This Paper}                 & \cmark     & \cmark     & \cmark     \\
    \bottomrule
  \end{tabular}
  }
	\vspace{0.1cm}
	\caption{Selection of publications related to this paper from the last five years.  All approaches cited above are based on code comment improvement using neural networks and the use of grounded theory in one way or another.  Column $G$ means the use of the grounded theory approach.  $L$ means LLM-based code summarization and comment generation.  $C$ means code comments generation and quality assessment.}
	\label{tab:screlated}
\end{table}

\subsection{Studies of Code Comment Quality}
Code comment quality is essential for ensuring the readability and maintainability of source code. Prior work has examined various aspects of assessing and evaluating comment quality, such as their adequacy \citep{arthur1989assessing}, content quality \citep{khamis2010automatic, steidl2013quality}, the co-evolution of comments and code \citep{fluri2009analyzing}, and the detection of inconsistent comments \citep{ratol2017detecting, wen2019large}. Several studies have proposed tools and techniques for the automatic assessment of comment quality \citep{khamis2010automatic, steidl2013quality, sun2016code, zhou2017analyzing}. In a recent systematic literature review, \cite{rani2023decade} cataloged 21 quality attributes (QAs), yet what constitutes a ``good'' comment remains a topic of debate. Our work addresses this gap by identifying the quality dimensions using the grounded theory approach, listed in Table~\ref{tab:codingtable}, and improving the comments along those quality dimensions.

\subsection{Code Summarization \& Comment Generation}

Source code summarization is the task of generating concise natural language descriptions of source code \citep{haiduc2010supporting}. It is a subset of code comment generation, where comments may describe larger codebases more broadly, while a ``summary'' specifically denotes a well-defined, high-level description of a particular code section (e.g., a method or class) \citep{su2024context}.

\textbf{Code Summarization:}
Source code summarization has been an active area of research in software engineering for over a decade \citep{su2024context, haiduc2010supporting}. Its goal is to abstract the functionality of code into a human-readable format, helping developers understand the purpose and behavior of code, particularly within large codebases \citep{su2024context}. Early efforts employed traditional rule-based approaches \citep{sridhara2011automatically, mcburney2014automatic, rodeghero2014improving}. More recently, neural network-based models \citep{bansal2023statement, leclair2019neural, leclair2020improved} and large language models (LLMs) such as OpenAI’s GPT-4 \citep{achiam2023gpt}, CodeLlama \citep{roziere2023code}, and CodeBERT \citep{feng2020codebert} have significantly advanced this task \citep{su2024context}. Despite these advancements, large models often fail to generate summaries that fully address categories such as “what,” “why,” “how-to-use,” “how-it-is-done,” “property,” and “others” \cite{chen2021my, sun2024source}. To address this limitation, we propose a reinforcement learning from human feedback (RLHF)~\cite{lee2023rlaif} method to generate code comments aligned with the quality axes defined in Table~\ref{tab:codingtable} by finetuning a smaller model, which is suitable for in-house deployment.

\textbf{Comment Generation:}
Complementing code summarization, automated comment generation focuses on improving or revising existing comments. Code comments are widely recognized as valuable aids for program comprehension \citep{sridhara2011automatically, hu2018deep}. Early studies primarily used template-based \citep{sridhara2010towards, mcburney2014automatic} and information retrieval (IR)-based methods \citep{haiduc2010supporting, wong2013autocomment} for comment generation. Since 2017, neural models \citep{liu2018neural, hu2018deep, leclair2019neural, wang2020reinforcement, wei2020retrieve, mu2022automatic, mu2023developer}  and large language models has advanced in the task to generate comments automatically. Recent work shows that large language models (LLMs) can leverage in-context learning to produce multi-intent comments \citep{geng2024large, sun2024source, haider2024prompting}. However, these models primarily aim to imitate human-written comments and often lack a nuanced understanding of comment quality \citep{rani2023decade, geng2024large, mu2023developer}. This misalignment motivates our work, which integrates comment quality considerations into the improvement process. Our goal is to bridge the gap between comment quality research and generation techniques, enabling method to generate code comments aligned with the quality axes defined in Table~\ref{tab:codingtable}.


\subsection{Grounded Theory in Software Engineering}
Grounded Theory (GT) is a methodology for inductively generating theory from data \citep{glasserandstratus}. Unlike approaches that test existing theories, GT focuses on theory generation \citep{ADOLPH20121269}, as introduced by \cite{glasserandstratus} in The Discovery of Grounded Theory. In software engineering, GT is commonly used to analyze qualitative data—such as developer interviews and surveys—to explain phenomena like collaboration or code evaluation \citep{COLEMAN2007654, Carver2004TheIO, stol2016grounded}. It is particularly effective for exploratory questions like “What’s going on here?” \citep{ADOLPH20121269}, allowing patterns and concepts to emerge from observations \citep{stol2016grounded}. Given the lack of established theory on code comment quality \citep{rani2023decade}, we adopt GT to derive a set of “quality axes” based on developer feedback and rationale. Among the three GT variants discussed by \cite{stol2016grounded}, we use the Straussian variant, which supports structured coding and allows guiding frameworks \citep{glasserandstratus}. Straussian GT involves three coding stages \citep{stol2016grounded, strauss1997grounded}:
\begin{itemize}
    \item \textbf{Open Coding} is the process of generating ``categories'' and examining their dimensional variations \citep{stol2016grounded}. As outlined in Sec. 4.1, we adopted an iterative approach in which human annotators independently proposed labels, followed by review sessions to merge and refine them into a finalized set.
    \vspace{3mm}
    \item \textbf{Axial Coding} involves reassembling the data after open coding by identifying relationships among the categories \citep{glasser1992grounded}. As described in Sec. 4.2, we conducted a series of review sessions to group the labels into broader themes.
    \vspace{3mm}
    \item \textbf{Selective/Subjective Coding} is the process of determining a central category that integrates all major categories \citep{corbin1998basics}. Following the method proposed by Strauss and Corbin, we identified an umbrella label that encapsulates the key factors underlying the rationales.
\end{itemize}

\section{Comment Quality Rationale Survey}\label{sec:study}

This section describes our survey to collect programmers' rationales about code comment quality differences.  In a nutshell, we show programmers Java methods and comments that describe those methods.  Then we ask programmers to rank the comments by their quality, then rewrite the ``better'' comment to improve it further.  We ask the programmers to write a rationale for their decisions about comment quality.  The purpose is to coax programmers to think critically about code comment quality and provide their subjective opinions.

\vspace{-0.5em}

\subsection{Research Method}\label{sec:studymethod}

\begin{figure}[h]
    \centering
    \includegraphics[width=\linewidth]{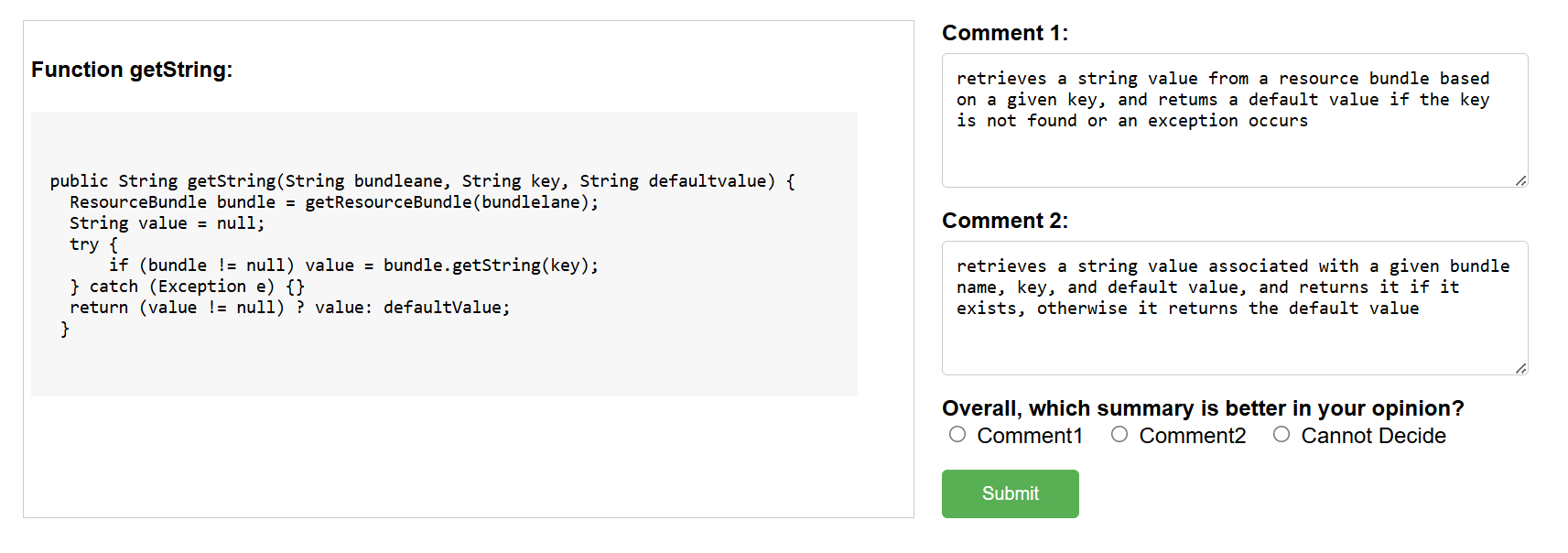}
    \caption{First page of the study interface.}\label{fig:firstPage}
\end{figure}

Our research method is a web-based survey with several pages.  The survey shows a series of Java methods.  There are two pages in the survey per method.  On the first page shown in Figure~\ref{fig:firstPage}, the human study participant sees a Java method's source code and two comments describing the method.  The participant then selects the comment that he or she prefers.  An option is provided for no preference, but this option is discouraged in the instructions.

\vspace{-1.5em}
\begin{figure}[h]
    \centering
    \includegraphics[width=\linewidth]{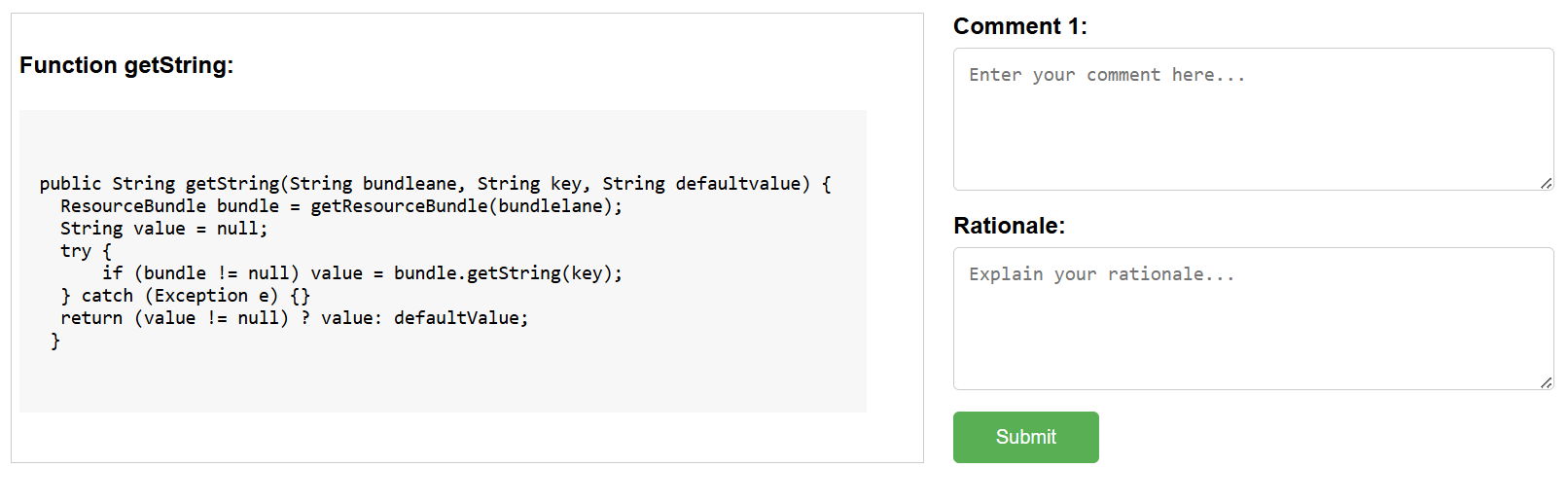}
    \caption{Second page of the study interface.}\label{fig:secondPage}
    
\end{figure}
After the participant indicates a preference (or none) for a comment, a second page shown in Figure~\ref{fig:secondPage} asks the participant to rewrite the comment that the participant chose (the survey chooses randomly if no preference is offered) to improve the comment further.  The participant is also asked to provide a brief rationale about the participant's decision about quality.  After finishing the task and clicking submit, the participant is shown another Java method.  The survey continues for a total of 35 Java methods.  We chose 35 methods because we found it takes approximately 90 minutes for participants to complete, after which various studies report a risk of fatigue bias~\citep{bansal2021project, mcburney2016automated, sridhara2010towards}.

\subsection{Source Code \& Comments Dataset}
\label{sec:dataset}
\vspace{-0.5em}
The origin of the Java methods and comments in our study is the dataset released by~\citep{su2023language} for source code summarization, which in turn is based on a dataset recommended by~\citep{leclair2019neural} and refined by~\citep{bansal2021project}.  The refinements addressed issues in datasets of code summaries raised by~\citep{shi2022we} and~\citep{allamanis2019adverse}.  We chose this dataset due to its extensive vetting and history in related research, as well as its large size and scope.  While it is tempting to view the dataset as ``one programming language'' (Java), it includes projects that represent a wide range of domains, programmers, and level of maintenance.

The dataset already included code summaries written by people for each Java method, but~\citep{su2024distilled} added comments generated by GPT-3.5 and a smaller language model that distilled GPT-3.5, for a total of three choices of comment for each method.  The dataset also includes a subset of around 8k Java methods that are even more carefully curated for quality and recommended as a test set in code summarization experiments.  We configured our survey to randomly select Java methods from the this 8k test set (again, up to 35 total).  Then, we randomly pick two of the three available summaries to display to the study participant for evaluation.  

Note that our goal is to collect programmers' rationales for code comment quality.  Given the subjective nature of these opinions, we do not consider it appropriate to enforce agreement.  Reasonable persons may disagree about preferences -- what we care about and analyze is the rationale.

\subsection{Participants}
\label{sec:participants}

We recruited 80 programmers from the Prolific platform~\citep{prolificWebsite} who were at least 25 years of age, were located in the United States or United Kingdom (for payment purposes), who had a university degree in Computer Science or Computer Engineering, and who self-report experience in Java. Ten programmers were assigned to each of the eight task: one comment quality rationale survey, and seven to evaluate the best comment along the axis. Remuneration was between \$20 and \$60/hour, depending on local market rates.

\begin{table}[H]
  \centering
  \caption{Labels from our Grounded Theory Process}
  \label{tab:codingtable}
  \resizebox{\columnwidth}{!}{%
    \begin{tabular}{ c p{0.5\columnwidth} c c c}
      \toprule
      \# 
        & \textbf{Description} 
        & \textbf{Label / Code} 
        & \textbf{Axial Code} 
        & \textbf{Subjective Code} \\
      \midrule
      1        
        & The preferred comment has additional information about the code logic of the method. The rationale may say things like: ``more detailed about how the method works'' or ``more detailed about how the algorithm is implemented.'' 
        & logical  
        & internalizing 
        & \\
      2        
        & The preferred comment has a specific variable name or data type, i.e., a very small additional that makes the comment less vague. The rationale may point out that the comment refers specifically to ``VariableName'' which makes it more detailed or clear.
        & precise    
        & internalizing 
        & \\
      3   
        & The preferred comment provides additional information that clarifies a feature of the method. For example, in a graphical program, ``fills a rectangular area'' versus ``fills a rectangular area \emph{with a specific color}.'' 
        & unambiguous   
        & internalizing 
        & \\
      4    
        & The preferred comment provides qualitatively more information in general without specifying information categorized in another label. The rationale may just say ``provides all required details'' or ``much more detailed overall.''  
        & exhaustive   
        & internalizing 
        & refocusing\\
        \\ \cdashline{1-4} \\
      5 
        & The preferred comment is considered better because it clarifies an error or exception type, reasons why an error or exception might occur, or examples of inputs that could cause a failure. 
        & troubleshooting 
        & externalizing 
        & \\
      6 
        & The preferred comment includes information to help understand why the method exists or what it is doing. For example, a method may play sounds, but a comment may contextualize the method by explaining that the purpose of the sound is for a game.  
        & contextualizing 
        & externalizing 
        & \\
      7      
        & The preferred comment is more concise than the non-preferred comment. Note that at times a comment may be more concise if it includes more focused information, even if the comment is longer in number of words.  
        & condensing 
        & externalizing 
        & \\ \\ \hdashline \\
      8            
        & Refers to rationales that are too vague to understand or contain non sequiturs, including potentially fraudulent survey responses. We remove these from further analysis.
        & toss  
        & n/a 
        & n/a \\
      \bottomrule
    \end{tabular}
  }
\end{table}
\vspace{-1.95em}
\subsection{Threats to Validity}


This study is subject to validity threats common to survey-based research in software engineering. Differences in the source code and comments, participant selection, or the survey interface could influence the results. To mitigate these risks, we sample from a large, vetted dataset and use a simple web interface to minimize complexity (unlike an IDE). While web surveys pose fraud risks~\citep{danilova2021you}, traditional countermeasures like challenge questions are increasingly ineffective due to tools like ChatGPT and Copilot~\citep{ghorbani2023autonomy}. Instead, we manually inspect and remove potentially fraudulent responses in the next section.


\section{Analysis of Comment Quality Rationale}\label{sec:analysis}

This section describes our qualitative analysis of the rationales we collected in the study in the previous section.  We use a Straussian grounded theory strategy described by~\cite{strauss1997grounded} and further elucidated for software engineering research in several articles~\citep{ADOLPH20121269, COLEMAN2007654, hoda2021socio, diaz2023applying} and especially by~\cite{stol2016grounded}. The Straussian grounded theory process is divided into three main steps: open coding, axial coding, and selective coding. Grounded theory approach has been effectively applied in software engineering to support collaborative qualitative analysis, for instance,~\cite{diaz2023applying}'s work on inter-rater reliability and agreement in collaborative grounded theory studies. While a complete review of grounded theory is beyond the scope of this paper, the following subsections summarize the key steps of the approach and describe our specific implementation.

\subsection{Open Coding}

Open coding is an approach in which annotators read subject artifacts and label each artifact with a single word or phrase.  The labels are defined by the annotators and are not from a predefined set.  The annotators may work independently, but then meet in review sessions to combine the labels.  The goal is to continue independent work followed by group meetings until a final set of labels is decided.  In our analysis, the subject artifacts are the rationales given in the survey in the previous study, and the annotators are the authors of this paper.  We pose the following research question and started our analysis individually and then in group review sessions:
\vspace{1em}
\begin{description}
    \item[\textbf{RQ1}] What are the factors that influence programmer opinions about code comment quality?
\end{description}
\vspace{1em}
Table~\ref{tab:codingtable} shows the outcome of our open coding process as a set of labels we attach to each rationale.  After finalizing the set of labels, we re-annotated each rationale with this final set.  In principle, a rationale could have more than one label, or a comment could be improved in multiple ways.  However, in our review sessions, we applied a dominant label to each rationale and resolved disagreements through discussion. This discussion was necessary because the highly subjective nature of rationales means multiple interpretations could be correct; the group review phase is intended to from a collective opinion about the most pertinent information.  Authors such as~\cite{delgado2019cohen} and~\cite{donker1993interpretation} argue against calculating Cohen's Kappa and other agreement metrics in similar scenarios, relying instead on collaborative discussions to resolve disagreement.

\begin{figure}[t]
    \centering
    \includegraphics[width=0.7\linewidth]{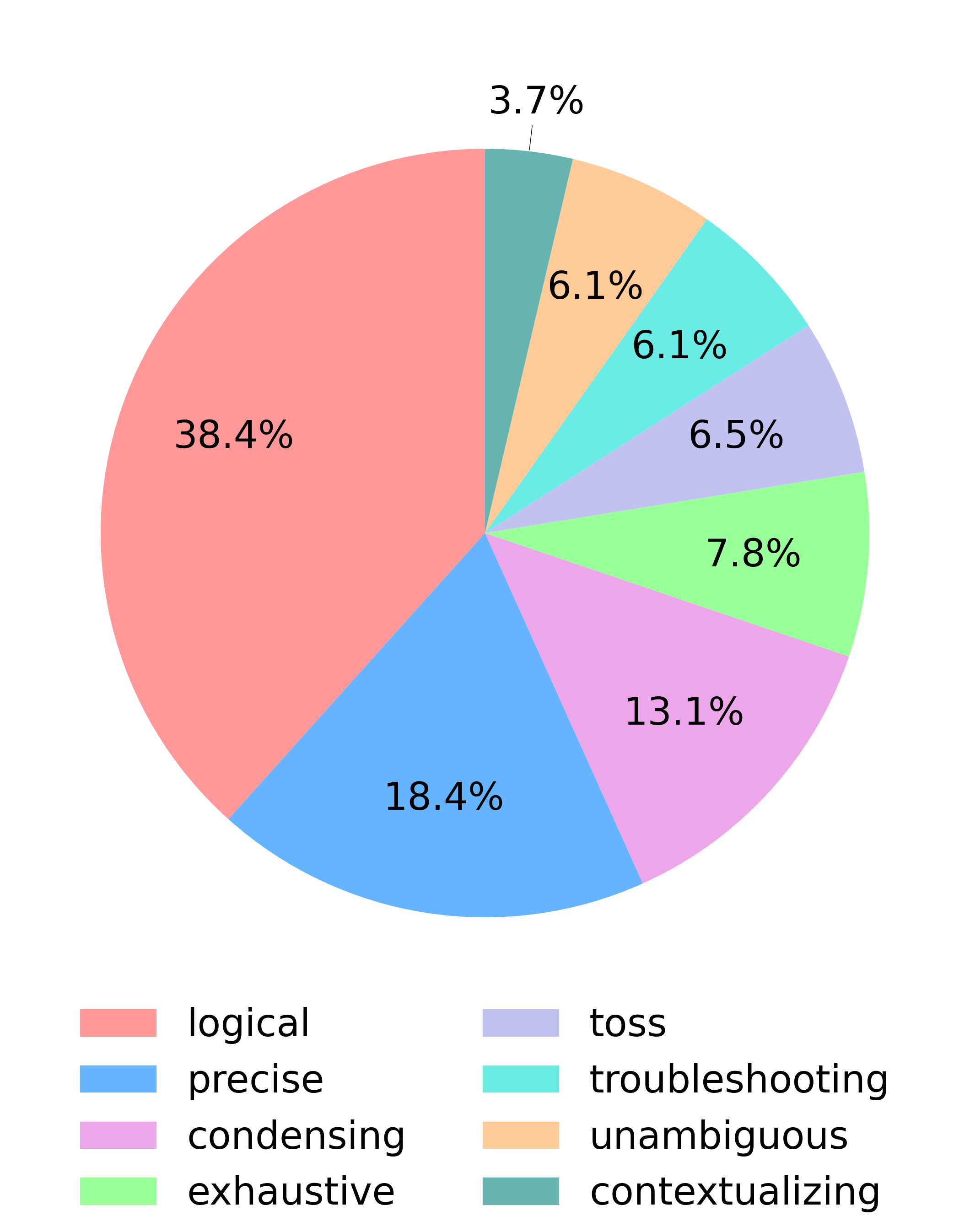}
    \caption{Distribution of tags from our annotation process.}
    \label{fig:labeldist}
\end{figure}

\subsection{Axial Coding}


The axial coding phase of grounded theory involves organizing the labels generated during open coding into higher-level categories that capture the relationships among the open codes. We carried out this process through a series of group review meetings. As a result, we identified two key themes as axial codes: \emph{internalizing} and \emph{externalizing}. The internalizing category encompasses labels referring to specific details within the method being described, whereas the externalizing category includes labels referring to information beyond or primarily affecting the method. For example, an internalizing improvement to a code description might address aspects such as variable names or the algorithms implemented within a method, such as the type of search algorithm employed. In contrast, an externalizing improvement might incorporate information explaining why the search is necessary in the program or identifying inputs from other components that could cause the search to fail. Note that we assigned the ``condensing'' label to the externalizing axial code group, as it consistently involved removing unnecessary details about the internal workings of the method in favor of emphasizing higher-level contextual information.

\subsection{Subjective Coding and Theory Summary}

The purpose of the subjective coding phase is to place the axial and open codes under a single umbrella code (the subjective code) that describes the key purpose behind the subject artifacts.  This subjective code serves as the lynchpin of the theory developed in the grounded theory process.  We found a subjective code of ``refocusing'' that crystallizes the key factors behind the rationales from our survey.  We define refocusing as changing the information in the source code comment to better align with one of several qualities that programmers need from the comments.  At a very high level, code comments improve when they discuss information along one of the qualities that we identify as labels during the open coding part of the grounded theory process.

The application of the labels is subjective and can vary depending on programmer preferences and the source code being described by the comment.  Consider the pie chart in Figure~\ref{fig:labeldist}.  Approximately 38\% of the labels are ``logical'' which on hand is not surprising because many comments that programmers write are focused on how an algorithm works or other implementation details.  But on the other hand, more than 60\% of the tags focus on other quality criteria, showing that programmers look for more than implementation details in comments.  The second-most common tag was ``precise'', with around 18\% of improvements related to very specific details such as the inclusion of a particular method name.  Note however that these tags are indicators of why study participants felt one summary was better than another -- they should not be interpreted as indicating what programmers need in an ideal case.  For example, the low percent for ``contextualizing'' (around 4\%) most likely indicates that contextual information is rare in summaries, rather than suggesting that programmers do not need contextual information.

\vspace{-0.5em}

\subsection{Examples}

To clarify the distinctions among labels, we present the illustrative examples.
Table~\ref{tab:rationaleExample} demonstrates how annotators assign the labels for rationales and Table~\ref{tab:labelsummary} illustrates the code, and summary along the quality axes.

Table \ref{tab:rationaleExample} illustrates the rationale labeling scheme. Rationale statements that address the behavior or functionality of the code are labeled as logical, while statements that emphasizes critical implementation details, such as excluding itsPreviewLocalizatorLayer when removing layers, and specifying that the upper activity layer is added only when an active project exists, making the comment more accurate and informative is labeled as precise. 

To further illustrate this categorization, Table \ref{tab:labelsummary} presents a Java code snippet along with corresponding summaries aligned with different quality axes. Each summary highlights a distinct aspect of the same code. The precise summary identifies the Ymsg object and emphasizes clarity in explanation, the logical summary describes the core functionality, the contextualize summary provides additional context—such as ``writing the appropriate byte array to the connection'', and the unambiguous summary is clear and free of ambiguity.

\begin{table}[ht]
\centering
\caption{Examples of rationale and label}\label{tab:rationaleExample}
\begin{tabular}{@{}lp{8cm}@{}}
\underline{\textbf{Label}} & \underline{\textbf{Rationale}} \\ \\
    {logical}         & Providing more clarity and specificity of the behavior of the code. \\ 
    {contextualizing}     & provide a more complete and specific description of the function's behavior \\ 
    {unambiguous}     & Providing more clarity about the storage manager and the nature of the loading process   \\ 
    {precise}         & Mention the exclusion of the itsPreviewLocalizatorLayer when removing layers from the camera. Additionally, I emphasize that the upper activity layer is added to the camera conditionally, specifically when an active project exists \\
\end{tabular}
\end{table}

\newsavebox{\codeboxexample}
\begin{lrbox}{\codeboxexample}
\lstset{
  language=Java,
  backgroundcolor=\color{backcolour},   
  commentstyle=\color{codegreen},
  keywordstyle=\color{magenta},
  numberstyle=\tiny\color{codegray},
  stringstyle=\color{codepurple},
  basicstyle=\ttfamily\footnotesize,
  breakatwhitespace=false,         
  breaklines=true,                 
  captionpos=b,                    
  keepspaces=true,                     
  numbersep=5pt,                  
  showspaces=false,                
  showstringspaces=false,
  showtabs=false,                  
  tabsize=2,
  frame=single,                    
  framesep=5pt,                    
  rulecolor=\color{black!30}       
}
\begin{lstlisting}
    public void rejectContact(String buddy, String message) throws IOException {
        YmsgPacket yp = new YmsgPacket(version, 0, Event.REJECTCONTACT, 0);
        yp.addData(new YmsgData(1, username));
        yp.addData(new YmsgData(7, buddy));
        yp.addData(new YmsgData(14, message));
        connection.write(yp.toByteArray());
    }
\end{lstlisting}
\end{lrbox}


\begin{table}[!ht]
\centering
\caption{Examples of label and summary}\label{tab:labelsummary}
\begin{tabular}{@{}l@{\hspace{1em}}p{0.73\linewidth}@{}}
\multicolumn{2}{@{}c@{}}{\usebox{\codeboxexample}} \\ \\ 
\underline{\textbf{Label}} & \underline{\textbf{Summary}} \\
\\
{contextualize} & Rejects a contact by transforming the chat message into an YmsgPacket, writing the Ymsg object, and ultimately writing the appropriate byte array to the connection. \\ 
{logical} & Rejects a contact by creating a YmsgPacket with specified parameters and writing it to the connection. \\ 
{unambiguous}         & rejects a contact by creating a YmsgPacket with specified parameters, adding data to it, and then writing it to the connection \\ 
{precise} & Rejects a contact by transforming the chat message into an YmsgPacket, writing the Ymsg object, and handling the response. \\ 
\end{tabular}
\end{table}
\begin{figure*}[ht]
    \centering
    \includegraphics[width=0.9\linewidth]{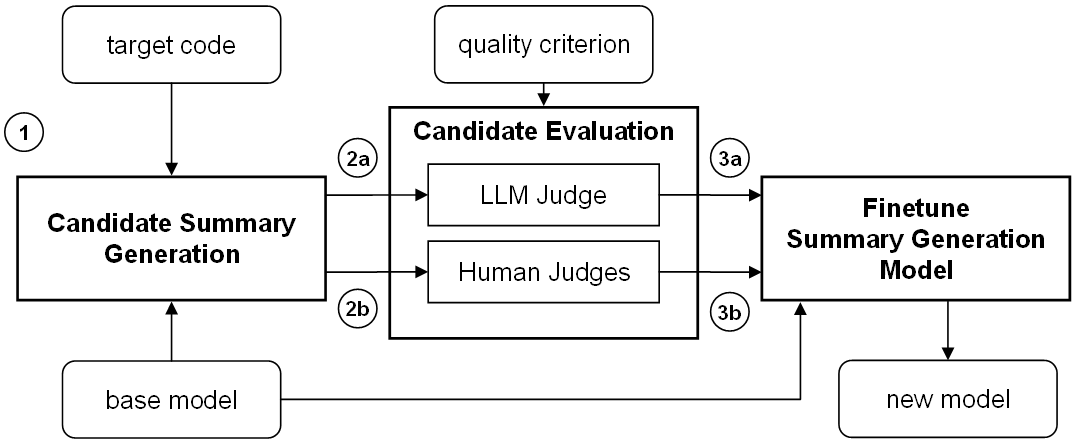}
    \caption{An overview of the three steps in our approach and their inputs.}
    \label{fig:overview}
\end{figure*}

\section{Improvement Approach}\label{sec:approach}

Our approach defines a systematic process for fine-tuning language models to generate code summaries optimized for one of the seven quality axes identified in the previous section. An overview of this process is presented in Figure~\ref{fig:overview}. The procedure involves three main steps: (1) using a pre-trained base model to generate multiple candidate summaries, (2) selecting the candidate that best aligns with a specified quality axis, and (3) fine-tuning the base model based on these selections. The result is a refined model capable of producing code summaries tailored to a specific quality dimension.


\subsection{Candidate Summary Generation}

The first step in our approach is to generate $n$ candidate summaries for each section of target code from a code dataset (Figure~\ref{fig:overview}, area 1). The purpose of this dataset is to train/fine-tune a model and is not the code for which summaries will be generated by a user of the new model. In our implementation, we used the Java dataset from~\cite{su2023language} that we also used in Section~\ref{sec:dataset}, though in principle many datasets are possible. We randomly selected 45k methods from training samples in that dataset. Then we used the base model to generate $n=4$ candidate summaries for each of the 45k methods.

\subsection{Evaluation of Candidate Summaries}

The second step is to pick one of the $n$ candidate summaries that is best along a given quality axis.  We had two parts to this selection.  First (Figure~\ref{fig:overview}, area 2a) we use a commercial LLM (GPT-4o in our implementation) to improve the summary in a ``reinforcement learning from AI feedback'' (RLAIF) strategy~\citep{lee2023rlaif, madaan2024self} which has been suggested in certain software engineering contexts~\citep{ahmed2024can}.  We create a prompt based on the description of each quality criterion from the previous section.  Each prompt has $n=4$ candidate summaries and the code for a Java method, as the example prompt shows in Figure~\ref{fig:prompt} (see our online appendix for all prompts, Section~\ref{sec:conclusion}).  The prompt asks the LLM to choose the best of the four options along a particular quality axis.  We use this prompt to select the best candidate for 44.5k of the 45k Java methods.  The result is one summary for each quality axis for each of the 44.5k methods.

The second part of the evaluation is to ask human experts to select the best summary along each quality axis (Figure~\ref{fig:overview}, area 2b) in a ``reinforcement learning from human feedback'' (RLHF) strategy~\citep{lee2023rlaif}.  We perform a survey for each quality axis.  The survey is similar to the one we used in Section~\ref{sec:studymethod}, except that it asks participants to choose a summary out of $n=4$ candidates and it provides instructions around each quality criterion (rather than overall personal preference).  We recruited ten participants for each quality criterion survey in the same manner as in Section~\ref{sec:participants}, but we restricted recruitment such that each person could only participate in one survey to reduce potential anchoring bias~\citep{furnham2011literature}.  Each survey showed 50 Java methods with summary candidates for a total of 500 samples. These methods were randomly selected from the pool of 45,000 methods for which candidate summaries were initially generated. Note that none of the 500 methods used in the surveys overlapped with the 44,500 methods used during the automated evaluation phase with the LLM. 

Basically our idea is to rely on the AI feedback for quantity and then human feedback for quality.  One can affordably obtain several thousand responses from the commercial LLM, but human subjective judgment is expensive and not practical at large scale.  Note that we are certainly not the first to use RLAIF or RLHF for model improvement, but instead demonstrate one way to combine them for code summarization.

\subsection{Finetuning the Summary Generation Model}

Our third step is to create a summary generation model for each quality criterion.  Our process is to start with an existing summary generation model as a base model, then fine-tune that model using the selections of candidates from the previous section.  We did finetuning using the 44.5k samples selected by the LLM judge first (Figure~\ref{fig:overview}, area 3a), then finetuned again using 400 of the 500 samples selected by the human judges (the other 100 we held out as a test set for the evaluation in the next section).  Our rationale for using the LLM selections first was to give the model a baseline with sufficient data sizes to generate code summaries along a particular quality criterion, even if the LLM selections are not ``perfect.''  However, we also used selections from human judges last, to modify the model's output with the best examples at the end of the process.


\begin{table}[b!]
\centering
\caption{Fine-tuning hyperparameters in our experiments.} \label{tab:hyperparams}
\begin{tabular}{@{}l|c|c@{}}
\toprule
\textbf{Parameter} & \textbf{JAM} & \textbf{CodeLlama} \\
\midrule
Epochs & 3 & 3 \\
Learning Rate & $3.0 \times 10^{-5}$ & $1.0 \times 10^{-4}$ \\
Batch Size & 4 & 2 \\
Gradient Accumulation Steps & 32 & 16 \\
LoRA $r$ & -- & 64 \\
LoRA $\alpha$ & -- & 16 \\
LoRA Dropout & -- & 0 \\
Quantization Type & -- & nf4 \\
Bits & -- & 4 \\
\bottomrule
\end{tabular}
\end{table}

\begin{figure}[H]  
\centering
\begin{minipage}{\textwidth}
\lstset{
  language=Java,
  backgroundcolor=\color{backcolour}, 
  basicstyle=\ttfamily\small,
  breakatwhitespace=false,         
  breaklines=true,                 
  captionpos=b,                    
  keepspaces=true,                     
  numbersep=5pt,                  
  showspaces=false,                
  showstringspaces=false,
  showtabs=false,                  
  tabsize=2,
  frame=single,                    
  framesep=5pt,                    
  rulecolor=\color{black!30}       
}
\begin{lstlisting}  
You are an expert Java developer with a focus on clear, concise, and accurate documentation.  Your task is to evaluate multiple summaries of a given Java method and select the one that best captures the method's core functionality, input specifications, primary operations, and error handling. When evaluating each summary, consider the following criteria:

1. Clarity and Precision: The ideal summary should be easy to understand and should precisely describe 
   what the method does. It should avoid vague language and accurately represent the method's purpose.
2. Appropriate Use of Variable Names and Data Types : Specific variable names or data types should be 
   included only when they enhance understanding or remove ambiguity. Unnecessary use of variable names or 
   data types should be avoided to keep the summary focused and clear.
3.  Coverage of Key Components :
   -  Functionality : The summary should capture the main purpose of the method without extraneous details.
   -  Input Specifications : Mention any key inputs, including data types and any constraints if needed.
   -  Primary Operations : Describe the main operations performed within the method, such as loops, conditional 
     checks, or transformations, if they relate to the purpose and help to clarify functionality of code.
   -  Error Management : If applicable, include any error handling or edge case management in a concise way.
4.  Avoid Redundancy : Try to exclude any descriptions or variable references that don't enhance the reader's understanding if possible. The summary should be streamlined to include essential information.

Here is the Java code and the summaries:
Java Code: {code}
Summaries:
1. {summaries[0]}
2. {summaries[1]}
3. {summaries[2]}
4. {summaries[3]}
5. Absolutely cannot decide.
Please select the most precise summary according to the criteria above:
- Use specific variable names or data types only when they clarify the functionality or remove ambiguity.
- Choose the clearest and most informative summary that best defines the overall functionality of the code.
If you cannot confidently choose the best summary, select option 5 ('Absolutely cannot decide') as a last resort. Return only the content of the summary you select, without adding extra information, numbers, or notes.

\end{lstlisting}
\caption{Prompt template we used for the first-pass LLM-based improvement for the ``precise'' quality axis.  We replace \{summaries[x]\} and \{code\} tags with a summary candidate ($0\leq x\leq n$, where $n$ is the number of candidates) and the method's Java source code.}
\label{fig:prompt}
\end{minipage}
\end{figure}

\subsection{Implementation Notes}

We built two implementations to demonstrate our approach.  The implementations are identical except for the base model.  One implementation uses \texttt{jam} model released by~\cite{su2024distilled}, and the second uses the 13B parameter version of \texttt{codellama} model by Meta~\citep{roziere2023code}.  We used the \texttt{jam} model because 1) it is small enough to be finetuned without LoRA~\citep{hu2022lora} or other ``parameter efficient'' procedures on a single GPU, which preserves data custody and reduces experimental variables, 2) was shown to still be large enough to produce code summaries on par with a commercial LLM~\citep{su2024distilled, bansal2021project}, and 3) is open-source with pretraining data published so we can control experimental variables.  We used the \texttt{codellama} model because 1) it represents a much larger model than jam (37x larger) designed for code intelligence tasks, 2) is still feasible to finetune on commercial hardware with LoRA~\citep{hu2022lora} procedures, and 3) is partially open-source.

Note we used the finetuning settings recommended by the creators of each model with the hyperparameters in Table~\ref{tab:hyperparams}.  For \texttt{jam},~\cite{su2024distilled} published recommendations specifically for code summarization tasks.  For \texttt{codellama}, we used default settings the creators recommended for several tasks because no specifics for code summarization were provided.

Our implementation platform comprised an Intel i9-10900X CPU, 256GB of system memory, and four NVidia A5000 GPUs, each with 24GB of video memory. The software environment included Python 3.10.12, PyTorch 2.0.1, TensorFlow 2.12.0, and CUDA 12.2.


\section{Evaluation}\label{sec:eval}

Our research objective in this evaluation is to determine if our approach from the previous section leads to models that generate summaries better adapted to each quality criterion.  By ``better adapted'' we mean summaries that are more like those that people selected as higher quality along a quality axis than summaries generated prior to finetuning with our approach.  Therefore we ask the research question (RQ):

\begin{description}
    \item[\textbf{RQ2}] Are summaries from models finetuned with our approach more like summaries selected for quality axes, than base models prior to finetuning, according to accepted code summary similarity metrics?
\end{description}

The rationale behind RQ2 is that the finetuning process should lead the model to generate summaries more like the ones the model saw during finetuning, and to evaluate the process we need to measure the similarity of the summaries from the finetuned and base models to a reference set.  The reference is the set of 100 selections by human participants in the survey in the previous section (recall there were 500 selections made by human participants, of which 400 were used for finetuning and 100 set aside for evaluation).  For this evaluation, we use the 100 set-aside Java methods as input to both finetuned and base models.  For each of those 100 methods we have samples selected as better along each quality axis.  We use text similarity metrics to compare the output of each model to these samples.  Because quality criterion differences may be subjective and involve nuances of language, we follow the suggestion of~\cite{haque2022semantic} and use semantic similarity measures (as opposed to n-gram comparison metrics).  Specifically, we calculate BERTScore (we report F1)~\citep{zhang2019bertscore} and USE~\citep{haque2022semantic}.  Following~\cite{roy2021reassessing} we also do Mann-Whitney tests to determine the statistical significance between scores for each model.  The Mann-Whitney test is appropriate because it is a paired, non-parametric test.  A paired test is appropriate because each model uses the same set of input Java methods.  A non-parametric test is appropriate because it is conservative with respect to distribution assumptions.

Threats to validity exist in this evaluation, though these are consistent with similar studies.  Key threats to validity include: 1) the set of Java methods, 2) the human participants who made the selections, and 3) the similarity metrics.  In theory, differences in these categories could lead to differences in the evaluation outcomes.  We mitigate the threat from the Java methods by choosing these methods from a large dataset that has been vetted in numerous previous papers~\citep{su2023language}.  We mitigate the threat from the human participants by having a total of 50 participants to reduce the impact of bias from any one person, and implementing quality control procedures described in Section~\ref{sec:studymethod}.  We mitigate the threat from metrics by using two well-studied metrics that have been specifically recommended for code summary evaluation~\citep{zhang2019bertscore, haque2022semantic}, and by using recommended statistical tests~\citep{mann1947test}.  A limitation to our study, though not necessarily a threat to validity, is that the reference set is selected as ``best'' along a particular category, though may or may not be the best in a theoretical sense.  In theory, a better summary could be written along a particular quality axis than the one chosen as best of $n$.  This is a limitation of all RLHF and RLAIF approaches and not unique to our study.


\section{Evaluation Results}\label{sec3}

Tables~\ref{tab:rq2jam}~and~\ref{tab:rq2codellama} show our evaluation results we use to answer RQ2.  Overall we observe strong improvement in terms of similarity to human-selected improved summaries as measured by automated metrics.  For \texttt{jam} we observe statistically-significant improvement in BERTScore for five of the seven quality axes and four of the seven for USE.  The mean level of improvement in BERTScore is 23\% and in USE is 3.6\%.  For \texttt{codellama} we observe statistically-significant improvement in both BERTScore and USE for all axes.  The mean level of improvement in BERTScore is 115\% and in USE is 30\%.  These results indicate higher similarity to human-selected improved summaries.

\begin{table}[ht]
\centering
\caption{Metrics and statistical test results for RQ2 for jam.}\label{tab:rq2jam}

\begin{tabular}{@{}lllll@{}}
\multicolumn{5}{c}{\textbf{(a) BERTScore (F1)}} \\[1ex]
~ & \textbf{Baseline} & \textbf{Finetuned} & \textbf{p-value} & \textbf{\% diff} \\ 
\midrule
\multicolumn{1}{l|}{unambiguous}     & 45.95    & 52.2      & 0.0155            & 13.60  \\
\multicolumn{1}{l|}{condensing}      & 47.34    & 47.02     & 0.64              & -0.68  \\
\multicolumn{1}{l|}{contextualizing} & 44.76    & 49.69     & 0.067             & 11.01  \\
\multicolumn{1}{l|}{exhaustive}      & 47.15    & 52.07     & 0.032             & 10.43  \\
\multicolumn{1}{l|}{logical}         & 42.93    & 49.66     & 0.0036            & 15.68  \\
\multicolumn{1}{l|}{precise}         & 47.64    & 53.7      & 0.0044            & 12.72  \\
\multicolumn{1}{l|}{troubleshooting} & 45.41    & 91.51     & \textless{}0.0001 & 101.52 \\

\end{tabular}

\vspace{3mm}


\begin{tabular}{@{}lllll@{}}
\multicolumn{5}{c}{\textbf{(b) USE}} \\[1ex]
~ & \textbf{Baseline} & \textbf{Finetuned} & \textbf{p-value} & \textbf{\% diff} \\
\midrule
\multicolumn{1}{l|}{unambiguous}     & 71.5175  & 76.0375   & 0.0187            & 6.32  \\
\multicolumn{1}{l|}{condensing}      & 71.3375  & 72.2375   & 0.48              & 1.26 \\
\multicolumn{1}{l|}{contextualizing} & 70.74    & 75.75     & 0.01              & 7.08  \\
\multicolumn{1}{l|}{exhaustive}      & 71.0975  & 74.26     & 0.08              & 4.45  \\
\multicolumn{1}{l|}{logical}         & 69.4775  & 72.915    & 0.06              & 4.95 \\
\multicolumn{1}{l|}{precise}         & 71.85    & 75.7175   & 0.035             & 5.38  \\
\multicolumn{1}{l|}{troubleshooting} & 69.72    & 71.8125   & \textless{}0.0001 & 3.00 \\
\end{tabular}
\end{table}

\begin{table}[htbp]
\centering
\caption{Metrics and statistical test results for RQ2 for CodeLLAMA}\label{tab:rq2codellama}

\begin{tabular}{@{}l|cccr@{}}
\multicolumn{5}{c}{\textbf{(a) BERTScore (F1)}} \\[1ex]
~ & \textbf{Baseline} & \textbf{Finetuned} & \textbf{p-value} & \textbf{\% diff} \\ 
\hline
Unambiguous     & 26.82 & 53.61 & $<$0.0001 & {99.89}  \\
Condensing      & 22.63 & 51.16 & $<$0.0001 & {126.07} \\
Contextualizing & 21.60 & 49.39 & $<$0.0001 & {128.66} \\
Exhaustive      & 25.89 & 39.31 & $<$0.0001 & {51.83}  \\
Logical         & 21.36 & 47.43 & $<$0.0001 & {122.05} \\
Precise         & 22.93 & 54.66 & $<$0.0001 & {138.38} \\
Troubleshooting & 21.35 & 50.42 & $<$0.0001 & {136.12} \\
\end{tabular}

\begin{tabular}{@{}l|cccr@{}}
\multicolumn{5}{c}{\textbf{(b) USE}} \\[1ex]
~ & \textbf{Baseline} & \textbf{Finetuned} & \textbf{p-value} & \textbf{\% diff} \\
\hline
Unambiguous     & 59.05 & 76.28 & $<$0.0001 & {29.18}  \\
Condensing      & 56.32 & 72.24 & $<$0.0001 & {28.27}  \\
Contextualizing & 56.85 & 72.37 & $<$0.0001 & {27.30}  \\
Exhaustive      & 57.57 & 66.91 & $<$0.0001 & {16.22}  \\
Logical         & 55.24 & 71.92 & $<$0.0001 & {30.20}  \\
Precise         & 54.31 & 76.54 & $<$0.0001 & {40.93}  \\
Troubleshooting & 53.93 & 74.81 & $<$0.0001 & {38.72}  \\
\end{tabular}
\vspace{-1em}
\end{table}

A few observations stand out among the results, however.  First, for \texttt{jam}, the axis with the worst performance was ``condensing'' for which we found no statistically-significant difference for either metric (BERTScore even reported a slightly negative difference).  We attribute this finding to the human-selected summaries shorter length for condensing: while people tended to select shorter summaries in terms of number of words for this quality axis, the finetuning process tended to yield summaries with different content and word usage than summaries with fewer (or more) words, on average.  A second observation is that the improvement levels for \texttt{jam} are much smaller than for \texttt{codellama}.  We attribute this finding to higher diversity in the initial summaries from \texttt{codellama}, meaning that the human-selected summaries are more different from the other initial summaries than in \texttt{jam}.  The higher diversity is likely due to the larger model size and larger dataset used to pretrain \texttt{codellama}.  The \texttt{jam} model builders focused exclusively on a curated dataset of Java, while \texttt{codellama} was trained using many languages and repositories.  One may view this as \texttt{codellama} having more ``headroom'' to improve both in terms of difference from initial summaries and in capacity to learn due to model size.

A third observation is that the improvement along ``troubleshooting'' for \texttt{jam} as measured by BERTScore is much higher than for USE or that we found for \texttt{codellama}.  We attribute this finding to \texttt{jam}'s tendency to copy the exception type name directly from the code into the summary, which was also prominent in the human-selected summaries for that quality axis.  BERTScore calculates similarity by aggregating semantic similarity of each token, so identical word usage leads to high scores.  In contrast, USE calculates semantic similarity of an entire sentence, so the contribution of a single identical word match is less important to the total score.  Therefore, when \texttt{jam} learned during finetuning to copy a particular token from the source code, it closely matches what humans selected in terms of word usage, which was detected more strongly in BERTScore than in USE.

\newsavebox{\codeboxjam}
\begin{lrbox}{\codeboxjam}

\lstset{
  language=Java,
  backgroundcolor=\color{backcolour},   
  commentstyle=\color{codegreen},
  keywordstyle=\color{magenta},
  numberstyle=\tiny\color{codegray},
  stringstyle=\color{codepurple},
  basicstyle=\ttfamily\footnotesize,
  breakatwhitespace=false,         
  breaklines=true,                 
  captionpos=b,                    
  keepspaces=true,                     
  numbersep=5pt,                  
  showspaces=false,                
  showstringspaces=false,
  showtabs=false,                  
  tabsize=2,
  frame=single,                    
  framesep=5pt,                    
  rulecolor=\color{black!30}       
}
\begin{lstlisting}
protected void notifyProcessingReasonlet(ReasonletInstance instance) {
    if(listeners == null) {
        return;
    }
    for(ReasonletContainerListener listener : listeners) {
        try {
            listener.processingReasonlet(instance);
        } catch(Exception e) {
            if( logger.isEnabledFor(Level.ERROR) ) {
                logger.error( String.format("error while notifying listener \%s for processingReasonlet", listener), e);
            }
        }
    }
}
\end{lstlisting}
\end{lrbox}

\begin{table}[!h]

\centering
\caption{Examples from jam model after finetuning} \label{tab:jamexample}
\vspace{1em}
\begin{tabular}{@{}lp{8cm}@{}}
\multicolumn{2}{c}{\usebox{\codeboxjam}} \\ \\
\end{tabular}
\begin{tabular}{@{}lp{8cm}@{}}
\textbf{Reference Summary}: & notifies all registered instances of processing reasonlet and handles any exceptions thrown during the process \\ \\
\textbf{Label} & \textbf{Improved Summary} \\
    {Condensing}         & notifies all registered ReasonletContainerListeners of a ProcessingReasonlet instance \\
    {Contextualize}         & notifies all listeners of processing a Reasonlet instance by invoking the processingReasonlet method on each listener \\ 
    {Logical}   & notifies all listeners of processing a Reasonlet instance by invoking their processingReasonlet method \\
\end{tabular}
    
\end{table}

\newsavebox{\codeboxllama}
\begin{lrbox}{\codeboxllama}
\lstset{
  language=Java,
  backgroundcolor=\color{backcolour},   
  commentstyle=\color{codegreen},
  keywordstyle=\color{magenta},
  numberstyle=\tiny\color{codegray},
  stringstyle=\color{codepurple},
  basicstyle=\ttfamily\footnotesize,
  breakatwhitespace=false,         
  breaklines=true,                 
  captionpos=b,                    
  keepspaces=true,                     
  numbersep=5pt,                  
  showspaces=false,                
  showstringspaces=false,
  showtabs=false,                  
  tabsize=2,
  frame=single,                    
  framesep=5pt,                    
  rulecolor=\color{black!30}       
}
\begin{lstlisting}
public Intervention peek(int i) {
        if (i < 1)
            throw new IndexOutOfBoundsException("Index " + i + " is lower than 1");
        
        if (i > this.interventionList.size())
            throw new IndexOutOfBoundsException("Index " + i + " is greater than stack size: " + this.interventionList.size());
        
        return interventionList.getFirst();
    }
\end{lstlisting}
\end{lrbox}

\begin{table}[!ht]
\centering
\caption{Examples from codellama model after finetuning} \label{tab:codellamaexample}
\vspace{1em}
\begin{tabular}{@{}lp{8cm}@{}}
\multicolumn{2}{c}{\usebox{\codeboxllama}} \\ \\
\end{tabular}
\begin{tabular}{@{}lp{8cm}@{}}
\textbf{Reference Summary}:     &   returns the first element in a stack of interprets, but throws an exception if the provided index is out of bounds \\  \\
\textbf{Label} & \textbf{Improved Summary} \\ 

{Logical}     & returns the first element of a list of intervention objects, throwing an exception if the index is out of bounds \\

{Precise}     & returns the first element of a list of intervention objects at a specified index, throwing an exception if the index is out of bounds \\

{Exhaustive}      & returns the first element of a linked list, throwing an exception if the given index is less than 1 or greater than or equal to the size of the list \\

{Troubleshoot}        & returns the first element of a stack of intervention objects, after checking if the index is within the bounds of the stack \\

\end{tabular}
    
\end{table}

Tables~\ref{tab:jamexample}~and~\ref{tab:codellamaexample} present illustrative examples, each comprising a Java code snippet, a reference summary selected by a human evaluator from a set of candidate summaries, and the corresponding improved summaries produced after fine-tuning. These examples demonstrate how the fine-tuned models enhance the clarity, specificity, and overall quality of code comments across various evaluation axes. Table~\ref{tab:jamexample} focuses on clarifying vague or ambiguous descriptions. For instance, the reference summary—``notifies all registered instances of processing reasonlet''—is vague and lacks sufficient context. The refined summaries address these limitations by being more concise and informative. Specifically, the ``Condensing'' axis promotes brevity while preserving meaning; ``Contextualize'' introduces the specific listener type and the method being invoked; and ``Logical'' ensures the comment clearly articulates the intended action and its target. Table~\ref{tab:codellamaexample} further illustrates improvements across multiple quality dimensions. The reference summary in this example vaguely refers to a ``stack of interprets,'' which is both unclear and likely inaccurate. In contrast, the refined summaries improve terminology and detail: the ``Logical'' and ``Precise'' axes enhance accuracy and specificity; ``Exhaustive'' introduces explicit exception-handling conditions; and ``Troubleshoot'' emphasizes boundary checks, resulting in more robust and informative comments.

\section{Conclusion}\label{sec:conclusion}

This paper advances the state-of-the-art in two key ways:

\begin{enumerate}
    \item We conduct a grounded theory analysis of differences in code summaries to determine seven ``quality axes'' along which people view improvements in code summaries.
    \item We present and evaluate a strategy for fine-tuning language models to automatically improve code summaries along these seven quality axes.
\end{enumerate}

The intellectual merit/novelty of this paper is in the empirical findings from our grounded theory analysis, in the combination of RLAIF and RLHF techniques for code summary improvements, and in demonstrating that these techniques result in summaries that improve along the quality axes we found during the grounded theory analysis.  We measured improvement in terms of similarity to human-selected summaries using two metrics (one focusing on word usage similarity and one focusing on overall sentence similarity) and applied to two language models.  Overall we found stronger improvement in \texttt{codellama} as opposed to \texttt{jam}, which may be expected considering \texttt{codellama}'s larger size and pretraining dataset.




%

\section*{Declarations}

\subsection*{Ethical approval}\vspace{-0.5em}  
This study was conducted in accordance with institutional guidelines and approved by the Institutional Review Board.

\subsection*{Informed consent}\vspace{-0.5em}  
We presented the informed consent on the first page of the survey website, and all participants agreed to the study.

\subsection*{Author Contributions}\vspace{-0.5em}  
All authors contributed to the study’s conception and design. Material preparation, data collection, and analysis were performed by Maria Dhakal, Chia-Yi Su, and Aakash Bansal. Maria Dhakal and Collin McMillan drafted the manuscript, and all authors reviewed and approved the final version.

\subsection*{Code / Data Availability Statement}\vspace{-0.5em}  
We release all data, scripts, survey materials, and results at \url{https://github.com/apc1-research/improving_summary}

\subsection*{Conflict of Interest}\vspace{-0.5em}  
The authors declare no conflict of interest.

\subsection*{Clinical trial number: not applicable.}
\newpage
\bibliographystyle{spbasic}      
\bibliography{main}

\end{document}